# Species tree estimation using Neighbor Joining

Joseph Rusinko[*,1] and Matthew McPartlon[1]

[1]Departments of Mathematics and Computer Science, Hobart and William Smith Colleges

[*]**Corresponding Author**: E-mail: rusinko@hws.edu

**Abstract**

Recent theoretical work has demonstrated that Neighbor Joining applied to concatenated DNA sequences is a statistically consistent method of species tree reconstruction. This brief note compares the accuracy of this approach to other popular statistically consistent species tree reconstruction algorithms including ASTRAL-II Neighbor Joining using average gene-tree internode distances (NJst) and SVD-Quartets+PAUP*, as well as concatenation using maximum likelihood (RaxML). We find that the faster Neighbor Joining, applied to concatenated sequences, is among the most effective of these methods for accurate species tree reconstruction.

## Introduction

Dasarathy et al. (2015) introduced METAL (**M**etric algorithm for **E**stimation of **T**rees based on **A**ggregation of **L**oci), a species tree reconstruction algorithm which applies distance-based algorithms such as Neighbor Joining (NJ; Saitou and Nei 1987) to distances computed from molecular sequences concatenated at each loci. In their paper, Dasarathy et al. (2015) showed that METAL is statistically consistent using molecular sequences concatenated at each loci when distance is calculated under the Jukes Cantor substitution model (Jukes and Cantor, 1969). The authors also state that these results can be extended to more general models of evolution such as the General Time Reversible Model (GTR; Tavaré, 1986) using spectral techniques.

In addition to this, Yoshida and Nei (2016) demonstrated that running Neighbor Joining on distances corresponding to the proportion of sites that differ between taxa (uncorrected p-distances) generally outperforms popular phylogenetic reconstruction approaches, including Maximum Likelihood (ML; Felsenstein 1981) and Bayesian (Ronquist and Huelsenbeck 2003). These results hold in reconstructing both compositional and noncompositional genes. This is somewhat surprising, as p-distances do not make any correction for multiple substitutions at the same site, substitution rate biases or differences in evolutionary rates among sites.

In this note we compare Neighbor Joining with distances computed on concatenated aligned sequences (METAL) to four prominent species tree inference methods. While the original METAL protocol assumes data generated under the Jukes Cantor model, we use the term METAL to refer to the generalized practice of concatenating gene sequences, computing pairwise distances assuming the concatenated sequences arose as a single gene under a standard evolutionary model (for instance GTR+G), and then constructing a species tree by applying Neighbor Joining to these pairwise distances. We provide empirical support that METAL can be applied to distances computed under more general substitution models such as GTR+Gamma (NJ+GTR+G), and demonstrate that applying Neighbor Joining to uncorrected p-distances (NJ+p) computed on sequences concatenated at each loci is also surprisingly effective as a species tree reconstruction algorithm.

## Background

Much of the traditional work in phylogenetic inference has focused on the reconstruction of gene trees. However, true species and gene trees can be incongruent due to the effect of evolutionary processes like incomplete lineage sorting (ILS; Maddison, 1997), gene duplication, gene loss, and horizontal gene transfer (Kingman 1982; Degnan and Rosenberg 2006). Of these processes, ILS is one of the most common sources of discordance between gene trees and species trees (Edwards, 2009) and is statistically modeled by the multi-species coalescent (Kingman, 1982). In response to this, coalescent-based methods, which attempt to explicitly account for ILS, have been developed. Examples of such approaches include ASTRAL-II (Mirarab and Warnow 2015), Neighbor Joining using average gene-tree internode distances (NJst; Liu and Yu 2011), and SVQ-Quartets+PAUP* (Chifman and Kubatko 2014). Assuming error free gene tree reconstruction, ASTRAL-II and NJst are statistically consistent under the multi-species coalescent model, meaning that they will converge to the true species tree with high probability as the number of loci and sites per locus for each taxa increase. SVD-Quartets+PAUP*, on the other hand is statistically consistent without reference to gene tree reconstruction.

In addition to coalescent-based methods for species tree reconstruction, alternative approaches such as distance-based methods can be used to reconstruct a phylogeny. In general, these methods attempt to fit a tree to a matrix of pairwise genetic distances rather than work with the original data directly (Felsenstein, 1988). Some examples of popular distance-based methods include Neighbor Joining, and UPGMA (Sokal and Michener, 1958). The METAL algorithm, described in the introduction, is another example of a distance-based approach to species tree reconstruction, however, this approach differs from traditional Neighbor Joining in two main ways. First, METAL uses distances calculated on aligned molecular sequences concatenated at each loci, rather than the more traditional approach based on comparing sequences obtained from taxa at a single genetic locus. In the case of METAL, Neighbor Joining is applied on observed differences in the concatenated collection of DNA sequences. Second, the original NJ algorithm was developed to reconstruct gene trees based on a single collection of DNA sequence data, whereas METAL is meant to reconstruct species trees.

To avoid confusion, we point out that NJst is a distance-based algorithm that is lexically similar to the previously mentioned Neighbor Joining methods however, the distances used by NJst are calculated from internode distances on estimated gene trees, rather than being inferred directly from the sequence data. We would also like to note that although NJ+GTR+G is statistically consistent[1] without reference to gene tree reconstruction, the same can not be said for NJ+p. We included NJ+p in our study based on the recent work of Yoshida and Nei (2016), which suggests that model fit gained by using a complex model such as GTR+G, can be offset by errors in modeling the various parameters necessary in applying these substitution models. Thus the simpler NJ+p model which does not make assumptions about the evolutionary process may still be effective.

**Material and Methods**

We compare the accuracy of species trees constructed using NJ+GTR+G and NJ+p to four prominent species tree reconstruction algorithms. The four additional algorithms included in this study are those examined by Chou et al. (2015) including ASTRAL-II, NJst, and SVD-Quartets+PAUP* which are statistically consistent under the multi-species coalescent model, as well as concatenation using maximum likelihood, specifically RAxML (Stamatakis, 2006). The study by Chou et al. compared these approaches on four simulated datasets, varying by level of ILS, and found that most often the best results were obtained using ASTRAL-II, even on the shortest gene sequence alignments (10 sites per locus), however RAxML, which is not statistically consistent under the multi-species coalescent (Roch and Steel, 2014), was the most accurate of all methods under low ILS conditions.

We compare the performance of NJ+GTR+G and NJ+p using METAL on the same four simulated 11-taxon datasets. Each dataset contains gene sequence alignments for the same 11 taxa. The level of ILS is reflected in the average topological distance between simulated gene trees and the true species tree. None of the datasets assume a strict molecular clock, and within each dataset, gene sequence alignments differ by the number of sites sampled per gene, and the number of genes per sequence. For additional information on simulation procedures, model specifications, and access to the data used in this study see (Chou et al. 2015, pp.3-4, 8-10).

---

[1] For more information on substitution models that are statistically consistent using METAL, see (Dasarathy et al., 2014, pp.2).

For each of these 11-taxon datasets, we compared the Normalized Robinson Foulds (RF) rates of species trees estimated using NJ-p to those obtained by Chou et al.(2015) for ASTRAL-II, NJst, SVD-Quartets+PAUP*, and RAxML as shown in Figure 1.

Figure 1: Comparison of the Normalized Robinson Foulds (RF) rates for estimated species trees on 11-taxon datasets

| Model | Sites/Gene | # Genes | ASTRAL-II | RAxML | SVD-Q | NJst | NJ+p | NJ+GTR+G |
|---|---|---|---|---|---|---|---|---|
| **M1 (15.5% ILS)** | 10 Sites | 100 | 0.24 | 0.14* | 0.25 | 0.26 | 0.17 | 0.16 |
| | | 500 | 0.16 | 0.05* | 0.17 | 0.18 | 0.12 | 0.08 |
| | | 1000 | 0.13 | 0.04* | 0.13 | 0.14 | 0.12 | 0.06 |
| | 200 Sites | 100 | 0.05 | 0.03* | 0.08 | 0.06 | 0.12 | 0.05 |
| | | 500 | 0.04 | 0.02* | 0.07 | 0.05 | 0.11 | 0.02* |
| | | 1000 | 0.03 | 0.01* | 0.06 | 0.04 | 0.11 | 0.03 |
| **M2 (38.3% ILS)** | 10 Sites | 100 | 0.27 | 0.22 | 0.33 | 0.31 | 0.20* | 0.22 |
| | | 500 | 0.16 | 0.10* | 0.23 | 0.17 | 0.13 | 0.10* |
| | | 1000 | 0.15 | 0.07* | 0.15 | 0.17 | 0.12 | 0.07* |
| | 200 Sites | 100 | 0.10 | 0.09 | 0.12 | 0.10 | 0.10 | 0.06* |
| | | 500 | 0.06 | 0.03 | 0.05 | 0.06 | 0.08 | 0.02* |
| | | 1000 | 0.06 | 0.03 | 0.05 | 0.05 | 0.08 | 0.02* |
| **M3 (66.3% ILS)** | 10 Sites | 100 | 0.37 | 0.37 | 0.42 | 0.47 | 0.30* | 0.31 |
| | | 500 | 0.20 | 0.20 | 0.27 | 0.33 | 0.15* | 0.15* |
| | | 1000 | 0.13 | 0.14 | 0.20 | 0.22 | 0.11 | 0.10* |
| | 200 Sites | 100 | 0.15* | 0.19 | 0.20 | 0.15* | 0.16 | 0.15* |
| | | 500 | 0.07 | 0.10 | 0.15 | 0.07 | 0.07 | 0.06* |
| | | 1000 | 0.08 | 0.07 | 0.11 | 0.07 | 0.07 | 0.06* |
| **M4 (85.0% ILS)** | 10 Sites | 100 | 0.66 | 0.68 | 0.75 | 0.78 | 0.56* | 0.57 |
| | | 500 | 0.37 | 0.44 | 0.46 | 0.57 | 0.32* | 0.32* |
| | | 1000 | 0.30 | 0.37 | 0.43 | 0.48 | 0.25* | 0.25* |
| | 200 Sites | 100 | 0.33* | 0.53 | 0.54 | 0.37 | 0.41 | 0.40 |
| | | 500 | 0.18* | 0.27 | 0.34 | 0.18* | 0.19 | 0.18* |
| | | 1000 | 0.14 | 0.22 | 0.28 | 0.17 | 0.14 | 0.12* |

*Note. – The RF rates for ASTRAL-II, RAxML, SVD-Quartets with PAUP\* (SVD-Q) and Neighbor Joining using average gene-tree internode distances (NJst) were taken from Figure 1 of the comparison study by Chou et al.(2015). A \* on a given RF rate for a given entry of the table denotes the lowest rate achieved by the five methods with respect to the model, number of genes, and number of sites per gene. Each model contains the same 11 taxa, and varies by level of ILS. For each model, 50 replicate sequence alignments were used to estimate 50 separate species trees for each combination of sites per gene (sites/gene) and total number of genes (# Genes). The RF distances between these 50 estimated species trees and the true species tree were averaged and expressed as a percentage to generate the RF rates shown.*

**Results**

Holding all else constant RF rates tend to increase as the level of ILS increases and decrease as the number of genes and or sites increase. Similar to the findings of Chou et al. (2014) for model M1 (lowest level of ILS), RAxML achieves the lowest RF rates. With this said, NJ+GTR+G greatly outperforms RAxML on higher ILS datasets, whereas NJ+GTR+G is only slightly worse than RAxML on the M1 model. We also see the greatest difference in RF rates between NJ+GTR+G and NJ+p under this model.

For higher levels of ILS (models M2-M4) NJ+GTR+G produces RF rates at least as low as the previously studied methods on all but one data set (M4. 200 sites, 100 genes). The margin of difference in RF rates between NJ+GTR+G and the methods studied by Chou et al. (2014) appear to decrease as the number of sites per gene, or number of genes increase. Furthermore, the difference in RF rates between NJ+p and NJ+GTR+G appear to be negligible under these higher ILS conditions.

Of the previously studied methods, ASTRAL-II is equally competitive with the METAL family of algorithms when the level of ILS is high (M3 and M4). NJst is also competitive under high ILS conditions when 200 sites per gene are sampled.

**Discussion and Conclusion**

The data indicates that, not only is the METAL protocol (here implemented as NJ+p or NJ+GTR+G) statistically consistent, it outperforms other statistically consistent methods under a wide range of ILS conditions. This is particularly interesting given that this distance-based approach is much faster, than more complex models and thus is easily scalable to much larger data sets.

The difference in error rates between species trees estimated with NJ+p and NJ+GTR+G and those estimated by Summary methods such as NJst and ASTRAL-II may arise from gene tree estimation errors. Summary methods require an alignment estimated on each locus and an estimated gene tree on each alignment, and then combine the resulting estimated gene trees into a species tree. Because summary methods are sensitive to gene tree estimation error (Roch and Warnow 2015), which is more likely to occur on short alignments, the observed trends may be attributed to this effect.

While NJ+p makes almost no assumptions (other than that the data should be treelike), it returns results comparable to many of the most sophisticated models. This suggests that NJ+p could serve as a baseline test for future species tree studies, in that any proposed reconstruction algorithm should have to outperform NJ+p in a simulation study under its own model conditions in order to justify the additional running time.

There are implementations of Neighbor Joining such as RAPIDNJ and NINJA which are free to download (http://birc.au.dk/Software/RapidNJ/) and have worst case running times $O(n^2)$ and average case running times $O(n^3)$ where $n$ is the number of taxa (Simonsen et al, 2011). The current implementation of RapidNJ to handle very large datasets (50,000+ taxa) efficiently on a normal desktop computer (Simonsen et al, 2011).

**Glossary**

**ILS**: Incomplete Lineage Sorting | **GTR:** General Time Reversible | **METAL**: Metric algorithm for Estimation of Trees based on Aggregation of Loci | **NJ:** Neighbor Joining | **NJ+p:** Neighbor Joining using p-distances under the METAL protocol | **NJ+GTR+G:** Neighbor Joining using the General Time

Reversible substitution model with gamma distributed rate variation among sites, under the METAL protocol | **RF:** Robinson Foulds